\documentclass[structabstract]{aa}  
\usepackage{graphicx}
\usepackage{array}
\usepackage{txfonts}
\usepackage{natbib}
\bibpunct{(}{)}{;}{a}{}{,}

\begin{document}

\title{Optimal binning of X-ray spectra and response matrix design}

\author{J.S. Kaastra\inst{1,2,3}
\and
J.A.M. Bleeker\inst{1,3}
}

\offprints{J.S. Kaastra}
\date{\today}

\institute{SRON Netherlands Institute for Space Research, Sorbonnelaan 2,
           3584 CA Utrecht, the Netherlands 
           \and
           Leiden Observatory, Leiden University, PO Box 9513, 
           2300 RA Leiden, the Netherlands
           \and
           Department of Physics and Astronomy, Universiteit Utrecht, 
           P.O. Box 80000, 3508 TA Utrecht, the Netherlands
         }
         
\abstract
{  }
{ A theoretical framework is developed to estimate the optimal binning of X-ray
spectra.} 
{ We derived expressions for the optimal bin size for model spectra as well as
for observed data using different levels of sophistication. }
{ It is shown that by taking into account both the number of photons in a given
spectral model bin and their average energy over the bin size, the number of
model energy bins and the size of the response matrix can be reduced by a factor
of $10-100$. The response matrix should then contain the response at the bin
centre as well as its derivative with respect to the incoming photon energy. We
provide practical guidelines for how to construct optimal energy grids as well
as how to structure the response matrix. A few examples are presented to
illustrate the present methods.} 
{ }

\keywords{Instrumentation: spectrographs -- Methods: data analysis 
-- X-rays: general }
\titlerunning{Optimal binning}
%\authorrunning{J.S. Kaastra}

\maketitle

\section{Introduction}

Until two decades ago X-ray spectra of cosmic X-ray sources were obtained using
instruments such as proportional counters and gas scintillation proportional
counters with moderate spectral resolution (typically 5--20~\%). With the
introduction of charge-coupled devices (CCDs) (ASCA, launch 1993) a major leap
in energy resolution has been achieved (up to 2~\%) and very high resolution has
become available through grating spectrometers, first on the Extreme Ultraviolet
Explorer (EUVE), and later on the Chandra and XMM-Newton observatories.

These X-ray spectra are usually analysed with a forwards folding technique.
First a spectral model appropriate for the observed source is chosen. This model
is convolved with the instrument response, which is represented usually by a
response matrix. The convolved spectrum is compared to the observed spectrum and
the parameters of the model are varied in a fitting procedure in order to obtain
the best solution.

This classical way of analysing X-ray spectra has been widely adopted and is
implemented, e.g. in spectral fitting packages such as XSPEC \citep{arnaud1996},
SHERPA \citep{freeman2001}, and SPEX \citep{kaastra1996}. However, the
application of standard concepts, such as a response matrix, is not at all
trivial for high-resolution instruments. For example, with the RGS of
XMM-Newton, the properly binned response matrix is 120 Megabytes in size,
counting only non-zero elements. Taking into account that usually data from both
RGS detectors and of two spectral orders are fit simultaneously, makes it at
best slow to handle even by most present day computer systems. Also the higher
spectral resolution considerably enhances the computation time needed to
evaluate the spectral models. Since the models applied to Chandra and XMM-Newton
data are much more complex than those applied to data from previous missions,
computational efficiency is important to take into consideration.

For these reasons we discuss here the optimal binning of both model spectra and
data. In this paper, we critically re-evaluate the concept of response matrices
and the way spectra are analysed. In fact, we conclude that it is necessary to
drop the classical concept of a matrix, and to use a modified approach instead.

The outline of this paper is as follows. We start with a more in-depth
discussion regarding the motivation for this work (Sect.~\ref{sect:motivation}).
In the following sections, we discuss the classical approach to spectral
modelling and its limitations (Sect.~\ref{sect:classical}), the optimal bin size
for model spectra (Sect.~\ref{sect:modelbinning}) and data
(Sect.~\ref{sect:data}) followed by a practical example
(Sect.~\ref{sect:example}). We then turn to the response matrix
(Sect.~\ref{sect:matrix}), its practical construction
(Sect.~\ref{sect:construct}), and briefly to the proposed file format
(Sect.~\ref{sect:formats}) before reaching our conclusions.

\section{Motivation for this work\label{sect:motivation}}

In this section we present in more depth the arguments leading to the proposed
binning of model spectra and observational data and our choice for the response
matrix design. We do this by addressing the following questions.

\subsection{Why not use straightforward deconvolution?}

In high-resolution optical spectra the instrumental broadening is often small
compared to the intrinsic line widths. In those cases it is common practice to
obtain the source spectrum by dividing the observed spectrum at each energy by
the nominal effective area (straightforward deconvolution).

Although straightforward deconvolution, due to its simplicity would seem to be
attractive for high-resolution X-ray spectroscopy, it fails in several
situations. For example, spectral orders may overlap as with the EUVE
spectrometers \citep{welsh1990} or the Chandra \citep{weisskopf1996} Low-Energy
Transmission Grating Spectrometer \citep[LETGS;][]{brinkman2000} when the HRC-S
detector is used. In these cases only careful instrument calibration in
combination with proper modelling of the short wavelength spectrum can help. In
case of the Reflection Grating Spectrometer \citep[RGS;][]{denherder2001} on
board XMM-Newton \citep{jansen2001} 30\% of all line flux is contained in broad
wings as a result of scattering on the mirror and gratings, and this flux is
practically impossible to recover by straightforward deconvolution. Finally,
high-resolution X-ray spectra often suffer from both severe line blending and
relatively large statistical errors because of low numbers of counts in some
parts of the spectrum. This renders the method unsuitable.

\subsection{Why are high-resolution spectra much more complex to handle than
low-resolution spectra?}

The enhanced spectral resolution and sensitivity of the current high-resolution
X-ray spectrometers require much more complex source models with a multitude of
free parameters as compared to the crude models that suffice for low-resolution
spectra to cover all relevant spectral details that can be exposed by the far
superior resolving power of state-of-the-art cosmic X-ray spectrometers. This
does not merely involve adding more lines to the old models with the same number
of parameters. Whereas investigators first fit single-temperature, solar
abundance spectra, now multi-temperature, free abundance models are to be
employed for stars and clusters of galaxies, among others. Models of AGN have
evolved from simple power laws with a Gaussian line profile occasionally
superimposed to complex continua spectra, including features due to reflection,
relativistic blurring, soft excesses, multiple absorbing photo-ionised outflow
components, low-ionisation emission contributions originating at large
distances, which yet again leads to much larger numbers of free parameters to be
accounted for.

\subsection{Why is it not possible to model complex spectra by the sum of a simple continuum plus
delta lines?}

Practically all X-ray spectra in the Universe are too complex to simulate with
simple continuum shapes with superimposed $\delta$-functions mimicking line
features when observed at high resolution. Apart from spectral lines, other
narrow features often found are: for example, radiative recombination continua,
absorption edges, Compton shoulders, or dust features. Moreover spectral lines
are too complex to be dealt with by delta functions, i.e. the astrophysics
deriving from Doppler broadening or natural broadening, or a combination thereof
(Voigt profiles) remains totally unaccounted for.

Even if one would model a line emission spectrum by the sum of $\delta$-lines,
however, a physical model connecting the line intensities is needed. The line
fluxes are mutually not independent and, as a consequence, many weak lines may
not be detected individually but added together they may give a detectable
signal.

Furthermore, except for the nearest stars, almost all X-ray sources are subject
to foreground interstellar absorption, also yielding narrow and sharp spectral
features such as absorption edges or absorption lines (for instance the
well-known \ion{O}{i} 1s--2p transition at 23.5~\AA).

As an example, Seyfert 1 galaxies show a range of emission lines with different
widths, often superimposed on each other and corresponding to the same
transition, from narrow-line region lines (width few 100 km\,s$^{-1}$),
intermediate width lines (1000 km\,s$^{-1}$), broad-line region lines (several
1000 up to tens of thousands km\,s$^{-1}$) up to relativistically broadened
lines. In such cases splitting in narrow- and broadband features is impossible,
as there is spectral structure at many different (Doppler) scales. As already
stated above, there is a need for physically relevant models, regrettably,
$\delta$-functions do not qualify for this.

In principle, the user could design an input model energy grid where there is
only substantially higher spectral resolution at the energies where the model
would predict narrow spectral features, for instance near foreground absorption
edges or strong sharp emission lines. However, the real source spectrum may
contain additional narrow features that are not anticipated by the user, and we
want to avoid recreating grids and matrices multiple times. For this reason, the
binning scheme proposed in this paper only depends on the properties of the
instrument and the observed spectrum in terms of counts per resolution element.

\subsection{Why not use precalculated model grids?}

Some spectral models are computationally intensive. We are aware of the method
of precalculating grids of models, such that the spectral fitting proceeds
faster. This is common practice with for instance the APEC model for collisional
plasmas as implemented in XSPEC or XSTAR models for photo-ionised plasmas. For a
limited set of parameters this is an acceptable solution. However, as has been
outlined above in many situations, astrophysical models require substantially
more free parameters than can be provided with grids for 2--4 free parameters.

As an example, for stellar spectra not only temperature but also density and the
UV radiation field (for He-like triplets) constitute important parameters.
However, the standard APEC implementation only has temperature and abundances as
free parameters.

Photo-ionised, warm absorber models for AGN depend non-linearly on the shape of
the ionising spectral energy distribution and on source parameters such as
abundances and turbulence. In realistic descriptions, this requires at least
five or more free parameters and is much too expensive to construct grids. Using
grids in such cases always implies less realistic models.

In addition, creating these grids offers efficiency gain in the fitting
procedure but shifts the computational burden to the grid creation. The
developers of APEC communicated to us that a full grid covering all relevant
temperatures requires of the order of a week computation time. A single XSTAR
model (1 spectrum) takes on average about 20 minutes to complete, hence even
modest grids may take a week or so to complete. So, also it is extremely useful 
to be able to limit the number of energy bins in those cases.

\subsection{Why not use brute force computing power?}

We note that for individual spectra extensive energy grids and large sizes of
response matrices can be handled with present-day computers, but the computation
time in folding the response matrix into the spectrum is substantial. This
holds, in particular, if this process has to be repeated thousands of times in
spectral fitting and error searches for models with many free parameters.

More importantly, in several cases users want to fit spectra of time-variable
sources taken at different epochs together, using spectral models where some
parameters are fixed between the observations while others are not. This
requires loading as many response matrices as number of observations, and we
have seen cases where such analyses simply cannot be carried out in this way.

There are also other cases where investigators have to rely on indirect methods
such as making maps of line centroids or equivalent widths, rather than full
spectral fitting. A striking example constitutes the Chandra CCD spectra of Cas
A. With 1 arcsec spatial resolution the remnant covers of order $10^5$
independent regions, which should ideally be fit together with a common
spectral model with position-dependent parameters.

\subsection{Some explicit examples of spectral analyses that require considerable
computation time}

A few typical examples may help to stipulate the relevance of our case regarding
required computing time.

\citet{mernier2015} analysed a cluster of galaxies with the XMM-Newton EPIC
detectors. These detectors still have a modest spectral resolution. The spectra
of three detectors in two observations with different background levels were fit
jointly. The model was a multi-temperature plasma model with about 50 free
parameters, including abundances and different instrumental and astrophysical
background components. Fits including error searches took about 5--10 hours on a
modern workstation. Additional time is needed to verify that fits do not get
stuck in local subminima. These authors are extending the work now to radial
profiles in clusters, incorporating eight annuli per cluster, on a sample of 40
clusters. Several months of cpu time using multiple workstations are needed to
complete this analysis.

\citet{kaastra2014b} studied the AGN Mrk~509. Chandra HETGS spectra were fit
jointly with archival XMM-Newton spectra. As a result of the amount of
information available by virtue of the high spectral resolution, the spectrum
was modelled using an absorption model that comprised the product of 28
components (four velocity components times seven ionisation components). With
the inclusion of free parameters for the modelling of the continuum and the
narrow and broad emission features, the model ended up with about 60 free
parameters. It took two weeks of computing time to obtain the final model
including the assessment of uncertainties on all parameters.

\citet{kaastra2014a} observed the Seyfert galaxy NGC 5548 in an obscured state.
Stacked RGS, pn, NuSTAR, and INTEGRAL data were analysed jointly. The model
(table S3 of that paper) had 35 free parameters and consisted of six`warm
absorber' components and two `obscured' components, giving a total of eight
multiplied transmission factors, which had to be determined iteratively in about
eight steps to reach full convergence. This is because the inner, obscurer
components affect the ionising spectral energy distribution for the outer, warm
absorber components. At each intermediate step a full grid of Cloudy models had
to be calculated to update the transmittance of the obscured nuclear spectrum.
The full computation also took several days in addition to months of trial to
set up the proper model and calculation scheme.

\section{Classical approach to spectral modelling and its
limitations\label{sect:classical}}

\subsection{Response matrices}

The spectrum of an X-ray source is given by its photon spectrum $f(E)$, a
function of the continuous variable $E$, the photon energy, and has units of,
e.g. photons\,m$^{-2}$\,s$^{-1}$\,keV$^{-1}$. To produce the predicted count
spectrum $s(E^\prime)$ measured by an instrument, $f(E)$ is convolved with the
instrument response $R(E^\prime,E)$ as follows:
\begin{equation}
s(E^\prime) = \int\limits_0^{\infty} R(E^\prime,E)f(E){\rm d}E,
\label{eqn:matrix_cont}
\end{equation}
where the data channel $E^\prime$ denotes the observed (see below) photon energy
and $s$ has units of counts\,s$^{-1}$\,keV$^{-1}$. The response function
$R(E^\prime,E)$ has the dimensions of an (effective) area, and can be given in,
e.g. m$^2$.

The variable $E^\prime$ is denoted as the observed photon energy, but in
practice it is often some electronic signal in the detector, the strength of
which usually cannot take arbitrary values, but can have only a limited set of
discrete values, for instance as a result of analogue to digital conversion
(ADC) in the processing electronics. A good example of this is the pulse-height
channel for a CCD detector. Alternatively, it may be the pixel number on an
imaging detector if gratings are used.

In almost all circumstances it is not possible to carry out the integration in
Eqn.~(\ref{eqn:matrix_cont}) analytically because of the complexity of both the
model spectrum and instrument response. For that reason, the model spectrum is
evaluated at a limited set of energies, corresponding to the same predefined set
of energies that is used for the instrument response $R(E^\prime,E)$. Then the
integration in eqn.~(\ref{eqn:matrix_cont}) is replaced by a summation. We call
this limited set of energies the model energy grid or for short the model grid.
For each bin $j$ of this model grid, we define a lower and upper bin boundary
$E_{1j}$ and $E_{2j}$, a bin centre $E_j=0.5(E_{1j}+E_{2j}),$ and a bin width
$\Delta E_j = E_{2j} - E_{1j}$.

Provided that the bin width of the model energy bins is sufficiently small
compared to the spectral resolution of the instrument, the summation
approximation to (\ref{eqn:matrix_cont}) is in general accurate. The response
function $R(E^\prime,E)$ has therefore been replaced by a response matrix
$R_{ij}$, where the first index $i$ denotes the data channel, and the second
index $j$ the model energy bin number.

We have explicitly 
\begin{equation}
S_i = \sum\limits_{j}^{} R_{ij}F_j,
\label{eqn:rmatrix}
\end{equation}
where now $S_i$ is the observed count rate in counts\,s$^{-1}$ for data channel
$i$ and $F_j$ is the model spectrum (photons\,m$^{-2}$\,s$^{-1}$) for model
energy bin $j$.

\subsection{Evaluation of the model spectrum}

The model spectrum $F_j$ can be evaluated in two ways. First, the model can be
evaluated at the bin centre $E_j$, essentially taking
\begin{equation}
F_j = f(E_j)\Delta E_j.
\label{eqn:Fcentral}
\end{equation}
This is appropriate for smooth continuum models such as blackbody radiation and
power laws. For line-like emission, it is more appropriate to integrate the line
flux within the bin analytically, taking
\begin{equation}
F_j = \int\limits_{\displaystyle{E_{1j}}}^{\displaystyle{E_{2j}}}
f(E){\rm d}E.
\label{eqn:Fintegral}
\end{equation}
Now a serious flaw occurs in most spectral analysis codes. The parameter $S_i$
is evaluated in a straightforward way using (\ref{eqn:rmatrix}). Hereby it is
tacitly assumed that all photons in the model bin $j$ have exactly the energy
$E_j$. This is necessary since all information on the energy distribution within
the bin is lost and $F_j$ is essentially only the total number of photons in the
bin. If the model bin width $\Delta E_j$ is sufficiently small this is no
problem, however, this is (most often) not the case.

An example of this is the standard response matrix for the ASCA SIS detector
\citep{tanaka1994}, in which the investigators used a uniform model grid with a
bin size of 10\,eV. At a photon energy of 1\,keV, the spectral resolution (full
width at half maximum; FWHM) of the instrument was about 50\,eV, hence the line
centroid of an isolated narrow line feature containing $N$ counts can be
determined with a statistical uncertainty of 50/(2.35$\sqrt{N})$\,eV. We assume
here for simplicity a Gaussian instrument response (FWHM is approximately
2.35$\sigma$, see Eq.~\ref{eqn:fwhm_gauss})). Thus, for a line with 400 counts
the line centroid can be determined with an accuracy of 1\,eV, ten times better
than the bin size of the model grid. If the true line centroid is close to the
boundary of the energy bin, there is a mismatch (shift) of 5\,eV between the
observed count spectrum and the predicted count spectrum at about the $5\sigma$
significance level. If there are more of these lines in the spectrum, it is
possible that a satisfactory fit (e.g. acceptable $\chi^2$ value) is never
obtained, even in cases where the true source spectrum is known and the
instrument is perfectly calibrated. The problem becomes even more worrisome if,
for example detailed line centroiding is performed to derive velocity fields.

A simple way to resolve these problems is just to increase the number of model
bins. This robust method always works, but at the expense of a lot of computing
time. For CCD-resolution spectra this is perhaps not a problem, but with the
increased spectral resolution and sensitivity of the grating spectrometers of
Chandra and XMM-Newton this becomes cumbersome.

For example, the LETGS spectrometer of Chandra \citep{brinkman2000} has a
spectral resolution (FWHM) between 0.040--0.076~\AA\ over the 1--175~\AA\ band.
A 85~ks observation of Capella \citep[obsid.~1248,][]{mewe2001,ness2001}
produced $N=$14\,000 counts in the \ion{Fe}{xvii} line at 15~\AA. Because line
centroids can be determined with a statistical accuracy of $\sigma/\sqrt{N}$,
this makes it necessary to have a model energy grid bin size $\sigma/\sqrt{N}$
of about 0.00014~\AA, corresponding to 278 bins per FWHM resolution element and
requiring 1.2 million model bins up to 175~\AA\ for a uniform wavelength grid.
Although this small bin size is less than the (thermal) width of the line,
larger model bin sizes would lead to a significant shift of the observed line
with respect to the predicted line profile and a corresponding significant
worsening of the goodness of fit. With future, more sensitive instruments like
Athena \citep{nandra2013} such concerns will become even more frequent.

Most of the computing time in thermal plasma models stems from the evaluation of
the continuum. The radiative recombination continuum has to be evaluated for all
energies, for all relevant ions and for all relevant electronic subshells of
each ion. On the other hand, the line power needs to be evaluated only once for
each line, regardless of the number of energy bins. Therefore the computing time
is approximately proportional to the number of bins. Therefore, a factor of
1\,000 increase in computing time is implied from the used number of ASCA-SIS
bins (1180) to the required number of bins for the LETGS Capella spectrum (1.2
million bins). 

Furthermore, because of the higher spectral resolution of grating spectrometers
compared with CCD detectors, more complex spectral models are needed to explain
the observed spectra, with more free parameters, which also leads to additional
computing time. Finally, the response matrices of instruments like the
XMM-Newton RGS become extremely large owing to extended scattering wings caused
by the gratings.

It is therefore important to keep the number of energy bins as small as possible
while maintaining the required accuracy for proper line centroiding. Fortunately
there is a more sophisticated way to evaluate the spectrum and convolve it with
the instrument response. Basically, if more information on the distribution of
photons within a bin is taken into account (like their average energy), energy
grids with broader bins (and hence with substantially fewer bins) can give
results as accurate as fine grids where all photons are assumed to be at the bin
center. 

\section{What is the optimal bin size for model
spectra?\label{sect:modelbinning}}

\subsection{Definition of the problem}

In the example of the LETGS spectrum of Capella we showed that to maintain full
accuracy for the strong and narrow emission lines, very small bin sizes are
required, which leads to more than a million grid points for the model
spectrum. 

Fortunately, there are several ways to improve the situation. Firstly, the
spectral resolution of most instruments is not constant, and one might adapt the
binning to the local resolution. Some X-ray instruments use this procedure. It
helps, but the improvement is not very good for instruments like LETGS with a
difference of only a factor of two in resolution from short to long wavelengths.

Secondly, the finest binning is needed near features with large numbers of
counts, for instance the strong spectral lines of Capella. One might therefore
adjust the binning according to the properties of the spectrum: narrow bins near
high-count regions and broader bins near low-count regions. As far as we know,
no X-ray mission utilises this procedure. Perhaps the main reason is that one
first needs to know the observed count spectrum before being able to create the
model energy grid for the response matrix. In most cases, the redistribution
matrix is either precalculated for all spectra, or is calculated on the fly for
individual spectra only to account for time-dependent instrumental features.

A third way to reduce the number of bins is to revisit the way model spectra are
calculated. As indicated before, in most standard procedures, the integrated
number of photons in a bin is calculated. In the convolution with the response
matrix it is then tacitly assumed that all photons of the bin have the same
energy, i.e. the energy $E_j$ of the bin centre. However, it could well be that
the bin contains only one spectral line; it makes a difference if the line is at
the lower or upper limit of the energy bin or at the bin centre. The proper
centroid of the line in the observed count spectrum, after convolution with the
response matrix, is only reproduced if in the model spectrum not only the number
of photons but also their average energy is accounted for. As we show (see
Fig.~\ref{fig:plotres}), accounting for the average energy of the photons within
a bin allows us to have an order of magnitude larger bin sizes.

In the procedure proposed here we combine all three options to reduce the number
of model energy bins: making use of the local spectral resolution, the strength
of the spectral features in number of photons $N,$ and accounting for the
average energies of the photons within bins.

\subsection{Binning the model spectrum}

\subsubsection{Zeroth order approximation}

We have seen in the previous subsection that the classical way to evaluate model
spectra is to calculate the total number of photons in each model energy bin,
and to act as if all flux is at the centre of the bin. In practice, this implies
that the true spectrum $f_0(E)$ within the bin is replaced by its zeroth order
approximation, $f_{1,0}(E)$, and is written as
\begin{equation}
f_{1,0}(E) = N\delta(E-E_j),
\label{eqn:f0}
\end{equation}
where $N$ is the total number of photons in the bin, $E_j$ the bin centre as
defined in the previous section, and $\delta$ is the Dirac delta-function.

We assume, for simplicity, that the instrument response is Gaussian, centred at
the true photon energy and with a standard deviation $\sigma$. Many instruments
have line spread functions (lsf) with Gaussian cores. For instrument responses
with extended wings (e.g. a Lorentz profile) the model binning is a less
important problem, since in the wings all spectral details are washed out, and
only the lsf core is important. For a Gaussian profile, the FWHM of the lsf is
written as
\begin{equation}
{\rm FWHM}\ = \sqrt{\ln(256)}\sigma \simeq 2.35\sigma.
\label{eqn:fwhm_gauss}
\end{equation}

How can we measure the error introduced with approximation (\ref{eqn:f0})? We
compare the cumulative probability distribution functions (cdf) of the true
spectrum and the approximation (\ref{eqn:f0}), which are both convolved with the
instrumental lsf. The approach is described in detail in
Sect.~\ref{sect:testing}, using a Kolmogorov-Smirnov test we derive the model
bin size $\Delta E$ for which in 97.5\% of all cases the approximation
(\ref{eqn:f0}) leads to the same conclusion as a test using the exact
distribution $f_0$ in tests of $f_0$ versus any alternative model $f_2$.

The approximation eqn.~(\ref{eqn:f0}) fails most seriously in the case that the
true spectrum within the bin is also a $\delta$-function, but located at the bin
boundary, at a distance $\Delta/2$ from the assumed line position at the bin
centre.

The maximum deviation $\delta_k$ of the absolute difference of both cumulative
distribution functions, $\delta_k = \vert F_0(x)-F_{1,0}(x) \vert$ (see also
eqn.~\ref{eqn:lambda_k}) occurs where $f_0(x)=f_{1,0}(x)$, as outlined at the
end of Sect.~\ref{sect:anap}. Because $f_{1,0}(x) = f_0(x-\Delta/2)$, we find
that the maximum occurs at $x=\Delta /4$. After some algebra we find that in
this case 
\begin{equation} 
\delta_k  = \delta_{k,0} = P(\Delta/4)-P(-\Delta/4) =
2P(\Delta/4) - 1  = \frac{\Delta}{2\sqrt{2\pi}\sigma}, \label{eqn:d0}
\end{equation} 
where $P$ is the cumulative normal distribution. This approximation holds in the
limit of $\Delta \ll \sigma$. Inserting (\ref{eqn:fwhm_gauss}) we find that the
bin size should be smaller than 
\begin{equation} \frac{\Delta}{{\rm FWHM}} <2.1289\ \lambda_k N^{-0.5}, 
\label{eqn:de0} 
\end{equation}
where the number $\lambda_k$ is defined by (\ref{eqn:epsilon_lambda}). Monte
Carlo results are presented in Sect.~\ref{sect:montecarlo}.

\subsubsection{First order approximation}

A further refinement can be reached as follows. Instead of putting all photons
at the bin centre, we can put them at their average energy. This first-order
approximation can be written as
\begin{equation}
f_{1,1}(E) = N\delta(E-E_a),
\label{eqn:f1}
\end{equation}
where $E_a$ is given by
\begin{equation}
E_a = \int\limits_{\displaystyle{E_{1j}}}^{\displaystyle{E_{2j}}}
f(E)E{\rm d}E.
\label{eqn:eq}
\end{equation}

For the worst case zeroth order approximation $f_{1,0}$ for $f_0$, namely the
case that $f_0$ is a narrow line at the bin boundary, the approximation
$f_{1,1}$ yields exact results. Thus, it constitutes a major improvement. In
fact, it is easy to see that the worst case for $f_{1,1}$ is a situation where
$f_0$ consists of two $\delta$-lines of equal strength: one at each bin
boundary. In that case, the width of the resulting count spectrum is broader
than $\sigma$. Again in the limit of small $\Delta$, it is easy to show that the
maximum error $\delta_{k,1}$ for $f_{1,1}$ to be used in the Kolmogorov-Smirnov
test is written as
\begin{equation}
\delta_{k,1} = \frac{1}{8\sqrt{2\pi e}} (\frac{\Delta}{\sigma})^2,
\label{eqn:d1}
\end{equation}
where $e$ is the base of the natural logarithms. Accordingly, the limiting bin
size for $f_{1,1}$ is expressed as
\begin{equation}
\frac{\Delta_1}{{\rm FWHM}} < 2.4418\ \lambda(R)^{0.5} N^{-0.25}.
\label{eqn:de1}
\end{equation}

It is seen immediately that for large $N$ the approximation $f_{1,1}$ requires a
significantly smaller number of bins than $f_{1,0}$.

\subsubsection{Second order approximation}

We can decrease the number of bins further by not only calculating the average
energy of the photons in the bin (the first moment of the photon distribution),
but also its variance (the second moment). In this case we approximate
\begin{equation}
f_{1,2}(E) = N\exp[(E-E_a)/2\tau^2)],
\label{eqn:f2}
\end{equation}
where $\tau$ is given by
\begin{equation}
\tau^2 = \int\limits_{\displaystyle{E_{1j}}}^{\displaystyle{E_{2j}}}
f(E)(E-E_a)^2{\rm d}E
\label{eqn:tau}
.\end{equation}

The resulting count spectrum is then simply Gaussian with the average value
centred at $E_a$ and the width slightly larger than the instrumental width
$\sigma$, namely $\sqrt{\sigma^2 + \tau^2}$.

The worst case again occurs for two $\delta$-lines at the opposite bin
boundaries, but now with unequal strength. It can be shown in the small bin
width limit that
\begin{equation}
\delta_{1,2} = \frac{1}{36\sqrt{6\pi}} (\frac{\Delta}{\sigma})^3
\label{eqn:d2}
\end{equation}
and that this maximum occurs for a line ratio of $3+\sqrt{3}:3-\sqrt{3}$. The
limiting bin size for $f_{1,2}$ is written as
\begin{equation}
\frac{\Delta_2}{{\rm FWHM}} < 2.2875\ \lambda(R)^{1/3} N^{-1/6}.
\label{eqn:de2}
\end{equation}

\subsection{Monte Carlo results\label{sect:montecarlo}}

In addition to the analytical approximations described above, we have performed
Monte Carlo calculations as outlined in Sect.~\ref{sect:mccalculations}. For our
approximations of order 0, 1, and 2, corresponding to accounting for the number
of photons only, both the number of photons and centroid and the number of
photons, centroid, and dispersion, respectively, we found the following
approximations:
\begin{equation}
\frac{\Delta}{{\rm FWHM}} = \min (1,y),\label{eqn:dey}
\end{equation}
with
\begin{eqnarray}
\mbox{order 0}\ &:&\ y=\frac{0.5707}{x^{1/2}}\Bigl( 1 + \frac{1.0}{x} \Bigr),
\label{eqn:app0}\\
\mbox{order 1}\ &:&\ y=\frac{1.404}{x^{1/4}}\Bigl( 1 + \frac{18}{x} \Bigr), 
\label{eqn:app1}\\
\mbox{order 2}\ &:&\ y=\frac{1.569}{x^{1/6}}\Bigl( 1 + \frac{1.14}{x^{1/3}}
\Bigr),\label{eqn:app2}
\end{eqnarray}
where\begin{eqnarray}
\mbox{order 0}\ &:&\ x = N_r(1+0.3\ln R), \label{eqn:appr0}\\
\mbox{order 1}\ &:&\ x = N_r(1+0.1\ln R), \label{eqn:appr1}\\
\mbox{order 2}\ &:&\ x = N_r(1+0.6\ln R). \label{eqn:appr2}
\end{eqnarray}
Here $N_r$ is the number of photons within a resolution element and $R$ the
number of resolution elements. The precise upper cut-off value of 1.0 is a
little arbitrary, but we adopted it here for simplicity as the same as for our
approximation of the data grid binning (see Sect.~\ref{sect:data}).

\begin{figure}[!htbp]
\resizebox{\hsize}{!}{\includegraphics[angle=-90]{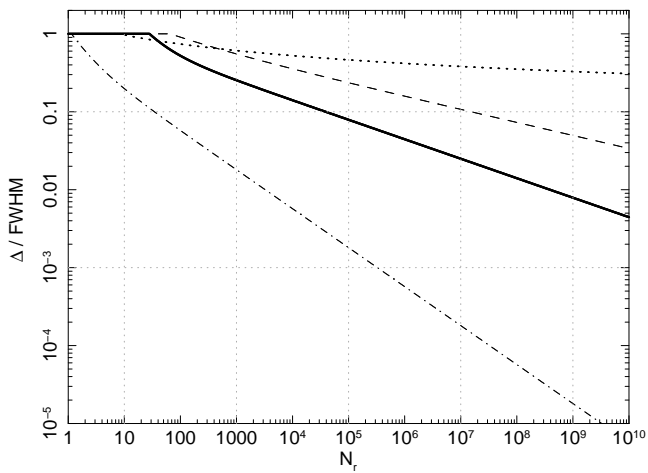}}
\caption{Optimal bin size $\Delta$ for model spectrum binning with a Gaussian
lsf, as a function of the number of counts per resolution element $N_r$.
Dash-dotted line: zeroth order approximation (\ref{eqn:app0}); thick solid line:
first order approximation (\ref{eqn:app1}); dashed line: second order
approximation (\ref{eqn:app2}). For comparison we also show as the dotted curve
the approximation (\ref{eqn:databin}) for the data grid binning. All results
shown are for $R=1$.}
\label{fig:plotres}
\end{figure}

We show these approximations in Fig.~\ref{fig:plotres}. The asymptotic behaviour
is as described by (\ref{eqn:de0}), (\ref{eqn:de1}), and (\ref{eqn:de2}), i.e.
proportional to $N_r^{-1/2}$, $N_r^{-1/4}$, and $N_r^{-1/6}$, respectively,
although the normalisations for the Monte Carlo results as compared to the
analytical approximations are higher by factors of 2.27, 1.63 and 1.38,
respectively. The difference is caused by the fact that the analytical
approximation assumed that the maximum of the Kolmogorov-Smirnov statistic
$D^\prime$ is reached at the $x$-value $x_m$ where the cumulative distributions
$F_0$ and $F_1$ reach their maximum distance. In practice, because of
statistical fluctuations the maximum for $D^\prime$ can also be reached for
other values of $x$ close to $x_m$, and this causes a somewhat more relaxed
constraint on the bin size for the Monte Carlo results.

\subsection{Which approximation do we choose?}

We now compare the different approximations $f_0$, $f_1$, and $f_2$ as derived
in the previous subsection. Fig.~\ref{fig:plotres} shows that the approximation
$f_1$ yields an order of magnitude or more improvement over the classical
approximation $f_0$. However, the approximation $f_2$ is only slightly better
than $f_1$. Moreover, the computational burden of approximation $f_2$ is large.
The evaluation of (\ref{eqn:tau}) is rather straightforward, although care
should be taken with single machine precision; first the average energy $E_a$
should be determined and then this value should be used in the estimation of
$\tau$. A more serious problem is that the width of the lsf should be adjusted
from $\sigma$ to $\sqrt{\sigma^2+\tau^2}$. If the lsf is a pure Gaussian, this
can be carried out analytically; however, for a slightly non-Gaussian lsf the
true lsf should be convolved in general numerically with a Gaussian of width
$\tau$ to obtain the effective lsf for the particular bin, and the computational
burden is very heavy. On the other hand, for $f_1$ only a shift in the lsf is
sufficient.

Therefore we recommend using the linear approximation $f_1$. The optimal bin
size is thus given by (\ref{eqn:dey}), (\ref{eqn:app1}), and (\ref{eqn:appr1}).

\subsection{The effective area}

Above we showed how the optimal model energy grid can be determined, taking into
account the possible presence of narrow spectral features, the number of
resolution elements, and the flux of the source. We also need to account for 
the energy dependence of the effective area, however. In the previous section,
we considered merely the effect of the spectral redistribution (rmf); here we
consider the effective area (arf).

If the effective area $A_j(E)$ would be a constant $A_j$ within a model bin $j$,
then for a photon flux $F_j$ in the bin the total count rate produced by this
bin would be simply $A_j F_j$. This approach is actually used in the classical
way of analysing spectra. In general $A_j(E)$ is not constant, however, and the
above approach is justified only when all photons of the model bin have the
energy of the bin centre. It is better to take into account not only the flux
$F_j$ but also the average energy $E_a$ of the photons within the model bin. 
This average energy $E_a$ is in general not equal to the bin centre $E_j$, and
hence we need to evaluate the effective area not at $E_j$ but at $E_a$.

We consider here the most natural first-order extension, namely the assumption
that the effective area within a model bin is a linear function of the energy.
For each model bin $j$, we develop the effective area $A(E)$ as a Taylor series
in $E-E_j$, which is written as
\begin{equation} 
A_j(E) = A(E_j) + A^\prime(E_j) (E-E_j) + \ldots.
\label{eqn:area_taylor}
\end{equation}
The maximum relative deviation $\epsilon_{\max}$ from this approximation occurs
when $E$ is at one of the bin boundaries. It is given by
\begin{equation}
\epsilon_{\max} = \frac{1}{8} (\Delta E)^2 A^{\prime\prime}(E_j)/A(E_j),
\label{eqn:epsilon_max}
\end{equation}
where $\Delta E$ is the model bin width. Therefore by using the approximation
(\ref{eqn:area_taylor}) we make at most a relative error in the effective area
given by (\ref{eqn:epsilon_max}). This can be translated directly into an error
in the predicted count rate by multiplying $\epsilon_{\max}$ by the photon flux
$F_j$. The relative error in the count rate is thus also given by
$\epsilon_{\max}$, which should be sufficiently small compared to the Poissonian
fluctuations $1/\sqrt{N_r}$ in the relevant range.

We can do this in a more formal way by finding for which value of $\epsilon$ a
test of the hypothesis that $N_r$ is drawn from a Poissonian distribution with
average value $\mu$ is effectively indistinguishable from tests that $N_r$ is
drawn from a distribution with mean $\mu(1+\epsilon)$. We have outlined such a
procedure in Sect.~\ref{sect:testing}, and in the limit of large $\mu$ we write
\begin{equation}
\mu + q_\alpha\sqrt{\mu} = 
 \mu (1+\epsilon ) + q_{k\alpha}\sqrt{\mu (1+\epsilon)},
 \label{eqn:epsilon_area}
\end{equation}
where $\alpha$ is the size of the test and $q_\alpha$ is given by
$G(q_\alpha)=1-\alpha$ with $G$ the cumulative normal probability distribution.
Further $k\alpha$ is the size of the test when we use the approximation
(\ref{eqn:area_taylor}). In the limit of small $\epsilon$, the solution of
(\ref{eqn:epsilon_area}) is given by
\begin{equation}
\epsilon \simeq p(\alpha,k) / \sqrt(N_r),
\end{equation}
where we have approximated $\mu$ by the observed value $N_r$, and where
$p(\alpha,k)=q_{\alpha}-q_{k\alpha}$. For $\alpha=0.025$ and $k=2,$ we obtain
$p(\alpha,k)=0.31511$.
Combining these results with (\ref{eqn:epsilon_max}), we obtain an expression
for the model bin size
\begin{equation}
\frac{\Delta}{{\rm FWHM}} = \Bigl( \frac{8Ap(\alpha,k)}{E_j^2 A^{\prime\prime}} 
\Bigr)^{0.5} 
\Bigl( \frac{E_j}{{\rm FWHM}} \Bigr)
N_r^{-0.25}.
\label{eqn:de1area}
\end{equation}

The bin width constraint derived here depends upon the dimensionless curvature
of the effective area ${A / E_j^2A^{\prime\prime}}$. In most parts of the energy
range this is a number of order unity or less. Since the second prefactor ${E_j
/ {\rm FWHM}}$ is by definition the resolution of the instrument, we see by
comparing (\ref{eqn:de1area}) with (\ref{eqn:de1}) that, in general,
(\ref{eqn:de1}) gives the most severe constraint upon the bin width. This is the
case unless either the resolution becomes of order unity, which happens, for
example for the Rosat \citep{truemper1982} PSPC \citep{pfefferman1987} detector
at low energies, or the effective area curvature becomes large, which may
happen, for example near the exponential cut-offs caused by filters.

Large effective area curvature due to the presence of exponential cut-offs is
usually not a problem, since these cut-offs also cause the count rate to be low
and hence weaken the binning requirements. Of course, discrete edges in the
effective area should always be avoided in the sense that edges should always
coincide with bin boundaries.

In practice, it is a little complicated to estimate from, for example a look-up table of
the effective area its curvature, although this is not impossible. As a
simplification for order of magnitude estimates, we use the case where
$A(E)=A_0{\rm e}^{bE}$ locally, which after differentiation yields
\begin{equation}
\sqrt{ \frac{8A}{ E_j^2A^{\prime\prime}} } = 
\sqrt{8}\, \frac{{\rm d}\ln E}{{\rm d}\ln A}.
\end{equation}
Inserting this into (\ref{eqn:de1area}), we obtain our recommended bin size,
as far as effective area curvature is concerned
\begin{equation}
\frac{\Delta}{{\rm FWHM}} = 1.5877\, 
 \Bigl( \frac{{\rm d}\ln E }{ {\rm d}\ln A} \Bigr)
 \Bigl( \frac{E_j}{{\rm FWHM}} \Bigr)
  N_r^{-0.25}.
\label{eqn:de1ar}
\end{equation}

\subsection{Final remarks}

In the previous two subsections we have given the constraints for determining
the optimal model energy grid. Combining both requirements (\ref{eqn:app1}) and
(\ref{eqn:de1ar}) we obtain the following optimal bin size:
\begin{equation}
\frac{\Delta}{{\rm FWHM}} = \frac{1}{1/w_1 + 1/w_a},
\label{eqn:de1tot}
\end{equation}
where $w_1$ and $w_a$ are the values of $\Delta$/FWHM as calculated using
(\ref{eqn:app1}) and (\ref{eqn:de1ar}), respectively.

This choice of model binning ensures that no significant errors are made either
due to inaccuracies in the model or the effective area for flux distributions
within the model bins that have $E_a\ne E_j$.

\section{\label{sect:data}Data binning}

\subsection{Introduction}

Most X-ray detectors count the individual photons and do not register the exact
energy value but a digitised version of it. Then a histogram is produced
containing the number of events as a function of the energy. The bin size of
these data channels ideally should not exceed the resolution of the instrument,
otherwise important information may be lost. On the other hand, if the bin size
is too small, one may have to deal with low statistics per data channel,
insufficient sensitivity of statistical tests or a large computational overhead
caused by the large number of data channels. Low numbers of counts per data
channel can be alleviated by using C-statistics \citep[often slightly modified
from the original definition by][]{cash1979}; computational overhead can be a
burden for complex models, but insufficient sensitivity of statistical tests can
lead to inefficient use of the information that is contained in a spectrum. We
illustrate this inefficient use of information in Appendix~\ref{sect:appc}. In
this section we derive the optimal bin size for the data channels. 

\subsection{The Shannon theorem}

The \citet{shannon1949} sampling theorem states the following: Let $f(x)$ be a
continuous signal. Let $g(\omega)$ be its Fourier transform, given by
\begin{equation}
g(\omega) = \int\limits_{-\infty}^{\infty}
f(x) {\rm e}^{\displaystyle{i\omega x}} {\rm d}x.
\label{eqn:fourier}
\end{equation}
If $g(\omega)=0$ for all $\vert \omega \vert > W$ for a given frequency $W$,
then $f(x)$ is band limited, and in that case Shannon has shown that
\begin{equation}
f(x) = f_s(x) \equiv \sum\limits_{n=-\infty}^{\infty}
f(n\Delta) \frac{\sin \pi(x/\Delta-n) }{ \pi(x/\Delta-n)}.
\label{eqn:shannon}
\end{equation}
In (\ref{eqn:shannon}), the bin size $\Delta=1/2W$. Thus, a band-limited signal
is completely determined by its values at an equally spaced grid with spacing
$\Delta$.

The above is used for continuous signals sampled at a discrete set of intervals
$x_i$. However, X-ray spectra are essentially a histogram of the number of
events as a function of channel number. We do not measure the signal at the data
channel boundaries, but we measure the sum (integral) of the signal between the
data channel boundaries. Hence for X-ray spectra it is more appropriate to study
the integral of $f(x)$ instead of $f(x)$ itself. 

We scale $f(x)$ to represent a true probability distribution. The cumulative
probability density distribution function $F(x)$ is written as
\begin{equation}
F(x) = \int\limits_{-\infty}^{x} f(y) {\rm d}y.
\label{eqn:cumdefinition}
\end{equation}
If we insert the Shannon reconstruction (\ref{eqn:shannon}) in
(\ref{eqn:cumdefinition}), after interchanging the integration and summation and
keeping into mind that we cannot evaluate $F(x)$ at all arbitrary points but
only at those grid points $m\Delta$ for integer $m$ where also $f_s$ is sampled,
we obtain
\begin{equation}
F_s(m\Delta) = \frac{\Delta }{ \pi}
\sum\limits_{n=-\infty}^{\infty} f(n\Delta)
\Bigl\{
\frac{\pi}{ 2} + {\rm Si}[\pi(m-n)]
\Bigr\}.
\label{eqn:shannon_int}
\end{equation}
The function ${\rm Si}(x)$ is the sine-integral as defined, for example in
\citet{abramowitz1965}. The expression (\ref{eqn:shannon_int}) for $F_s$ equals
$F$ if $f(x)$ is band limited. In that case at the grid points $F$ is completely
determined by the value of $f$ at the grid points. By inverting this relation,
one could express $f$ at the grid points as a unique linear combination of the
$F$-values at the grid. Since Shannon's theorem states that $f(x)$ for arbitrary
$x$ is determined completely by the $f$-values at the grid, we infer that $f(x)$
can be completely reconstructed from the discrete set of $F$-values. And then,
by integrating this reconstructed $f(x)$, $F(x)$ is also determined.

We conclude that $F(x)$ is also completely determined by the set of discrete
values $F(m\Delta)$ at $x=m\Delta$ for integer values of $m$, provided that
$f(x)$ is band limited.

For non-band-limited responses, we use (\ref{eqn:shannon_int}) to approximate
the true cumulative distribution function at the energy grid. In doing this, a
small error is made. The errors can be calculated easily by comparing
$F_s(m\Delta)$ with the true $F(m\Delta)$ values. The binning $\Delta$ is
sufficient if the sampling errors are sufficiently small compared with the
Poissonian noise. We elaborate on what is sufficient below.

\subsection{Optimal binning of data}

\begin{figure}[!htbp]
\resizebox{\hsize}{!}{\includegraphics[angle=-90]{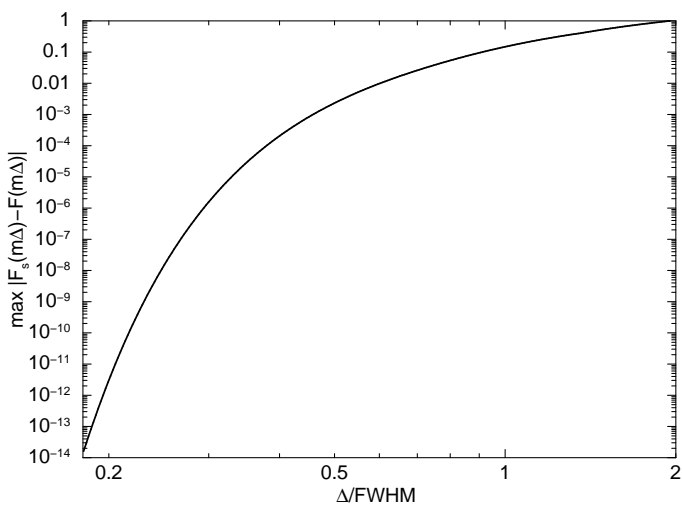}}
\caption{Maximum difference for the cumulative distribution function of a
Gaussian and its Shannon approximation, as a function of the bin size $\Delta$.
The maximum is also taken over all phases of the energy grid with respect to the
centre of the Gaussian.}
\label{fig:sagaus}
\end{figure}

\begin{figure}[!htbp]
\resizebox{\hsize}{!}{\includegraphics[angle=-90]{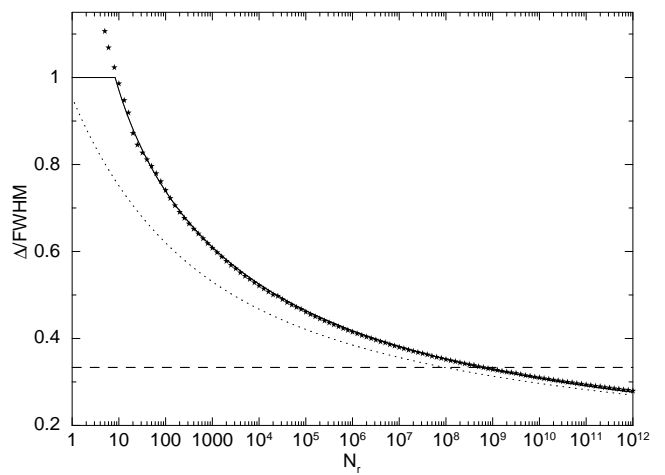}}
\caption{Optimal bin size $\Delta$ for data binning with a Gaussian lsf as a
function of the number of counts per resolution element $N_r$. Dotted curve:
analytical approximation using Sect.~\ref{sect:anap}; stars: results of Monte
Carlo calculation; solid line: our finally adopted bin size
Eq.~(\ref{eqn:databin}) based on a fit to our Monte Carlo results for large
values of $N_r$; dashed line: commonly adopted bin size 1/3~FWHM.}
\label{fig:databin}
\end{figure}

We apply the theory outlined above to the case of a Gaussian redistribution
function. We first determine the maximum difference $\delta_k$ between the
cumulative Gaussian distribution function $F_0(x)$ and its Shannon approximation
$F_1(x)=F_s(x)$, as defined by (\ref{eqn:lambda_k}). We show this quantity in
Fig.~\ref{fig:sagaus} as a function of the bin size $\Delta$. It is seen that
the Shannon approximation converges quickly to the true distribution for
decreasing values $\Delta$. Using $\delta_k=\lambda_k/\sqrt{N}$ (see
(\ref{eqn:epsilon_lambda})), for any given value of $N$ and $\lambda_k=0.122$ we
can invert the relation and find the optimal bin size $\Delta$. This is plotted
in Fig.~\ref{fig:databin}. We have also verified this with the Monte Carlo
method described before. 

It is seen that the Monte Carlo results indeed converge to the analytical
solution for large $N_r$. For $N_r=10$, the difference is 20\%. We have made a
simple approximation to our Monte Carlo results (shown as the solid line in
Fig.~\ref{fig:databin}), where we cut it off to a constant value for $N_r$ of
less than about 2. Both the Monte Carlo results and the analytical approximation
break down at this small number of counts. In fact, for a small number of counts
binning to about 1 FWHM is sufficient.

We also determined the dependence on the number of resolution elements $R$ for
values of $R$ of 1, 10, 100, 1000, and 10000. In general, the results even for
$R=10^4$ are not too different from the $R=1$ case. We found a very simple
approximation for the dependence on $R$ that is well described by the following
function:
\begin{equation}
\frac{\Delta}{\mathrm{FWHM}} = \left\{
\begin{array}{ll}
1 & \mbox{\ \ if $x\leq 2.119$};\\
\strut & \strut \\
\displaystyle{\frac{0.08 + 7.0/ x + 1.8 /x^2}{1 + 5.9 /x}} 
   & \mbox{\ \ if $x>2.119$},
\end{array} \right.
\label{eqn:databin}
\end{equation}
with
\begin{equation}
x \equiv \ln [ N_r (1+0.20 \ln R) ].
\label{eqn:gausdelta}
\end{equation}

We note that for digitisation of analogue electronic signals that represent
spectral information, a binning ($=$channel) criterion of 1/3~FWHM or better is
often adopted \citep[e.g.][]{davelaar1979}, since one does not know a priori the
level of significance of the measured signal (see the dashed line in
Fig.~\ref{fig:databin}).

\subsection{Final remarks\label{sect:datafinal}}

We have estimated conservative upper bounds for the required data bin size. In
the case of multiple resolution elements, we have determined the bounds for the
worst case phase of the grid with respect to the data. In practice, it is not
likely that all resolution elements would have the worst possible alignment.
However, we recommend using the conservative upper limits as given by
(\ref{eqn:gausdelta}).

Another issue is the determination of $N_r$, the number of events within the
resolution element. We have argued that an upper limit to $N_r$ can be obtained
by counting the number of events within one FWHM and multiplying it by the ratio
of the total area under the lsf (should be equal to 1) to the area under the lsf
within one FWHM (should be smaller than 1). For the Gaussian lsf, this ratio
equals 1.314; for other lsfs,it is better to determine the ratio numerically and
not to use the Gaussian value 1.314.

For the Gaussian lsf, the resolution depends only weakly upon the number of
counts $N_r$ (see Fig.~\ref{fig:databin}). For low count rate parts of the
spectrum, the binning rule of 1/3 FWHM usually is too conservative.

Finally, in some cases the lsf may contain more than one component. For example,
the grating spectra have higher order contributions. Other instruments have
escape peaks or fluorescent contributions. In general it is advisable to
determine the bin size for each of these components individually and simply take
the smallest bin size as the final one.

\section{A practical example\label{sect:example}}

In order to demonstrate the benefits of the proposed binning schemes, We have
applied them to the 85~ks Chandra LETGS observation of Capella mentioned
earlier. The spectrum spans the 0.86--175.55~\AA\ range with a spectral
resolution (FWHM) ranging between 0.040--0.076~\AA. In Table~\ref{tab:capella}
we show the number of data bins, model bins, and response matrix elements for
different rebinning schemes of this spectrum. For simplicity we consider only
the positive and negative first-order spectrum, and that the non-zero elements
of the response matrix span a full range of four times the instrumental FWHM. We
apply the algorithms as described in this paper to generate the data and model
energy grids.

The highest resolution of 0.040~\AA\ occurs at short wavelengths, and the
strongest line is the \ion{Fe}{xvii} blend at 17.06 and 17.10~\AA, with a
maximum number of counts $N_r$ of 15\,000 counts. 

The first case we consider (case A afterwards) is that we adopt a data and model
grid with constant step size for the full range. Accounting for the maximum
$N_r$ value, we have a data bin size of 0.02~\AA\ and a model bin size of
0.14~m\AA.

Case B is similar to case A except that we account for the variable resolution
of the instrument. This only decreases the number of bins by a factor of 1.375.

In case C we account for the number of counts in each resolution element, but we
still keep the ''classical'' approach of putting all photons at the centre of
the model bins. The number of resolution elements $R$ in the spectrum is 3110.

In case D we account for the average energy of the photons within a bin. The
number of model bins and response elements needed drops by more than an order of
magnitude for this case, as compared to case C.

\begin{table}[!htbp]
\caption{Number of data bins, model bins, and response matrix elements for
different rebinning schemes of the Chandra LETGS spectrum of Capella.}
\label{tab:capella}
\centerline{
\begin{tabular}{lrrr}
\hline\hline
Binning & Data bins & Model bins & Response \\
        &           &            & elements \\
\hline
A & $8.52\times 10^3$ & $1.26\times 10^6$ & $1.01\times 10^7$ \\
B & $6.20\times 10^3$ & $9.14\times 10^5$ & $7.31\times 10^6$ \\
C & $5.12\times 10^3$ & $1.23\times 10^5$ & $9.84\times 10^5$ \\
D & $5.12\times 10^3$ & $8.21\times 10^3$ & $6.57\times 10^4$ \\
\hline\noalign{\smallskip}
\end{tabular}
}
\end{table}

\section{The response matrix\label{sect:matrix}}

We have shown in the previous sections how the optimal model and data energy
grids can be constructed. We have proposed that for the evaluation of the model
spectrum both the number of photons in each bin as well as their average energy
should be determined. We now determine how this impacts the concept of the
response matrix.

In order to acquire high accuracy, we need to convolve the model spectrum for
the bin, approximated as a $\delta$-function centred around $E_a$, with the
instrument response. In most cases we cannot do this convolution analytically,
so we have to make approximations. From our expressions for the observed count
spectrum $s(E^\prime)$, eqns.~(\ref{eqn:matrix_cont}) and (\ref{eqn:rmatrix}),
it can be easily derived that the number of counts or count rate for data
channel $i$ is given by
\begin{equation}
S_i = \int\limits_{\displaystyle{E^\prime_{i1}}}^{\displaystyle{E^\prime_{i2}}}
    {\rm d}E^\prime \int\limits_{0}^{\infty}  R(E^\prime,E)f(E) {\rm d}E,
\label{eqn:countint}
\end{equation}
where, as before, $E^\prime_{i1}$ and $E^\prime_{i2}$ are the formal channel
limits for data channel $i$ and $S_i$ is the observed count rate in counts/s for
data channel $i$. Interchanging the order of the integrations and defining the
mono-energetic response for data channel $i$ by $\tilde R_i(E)$ as follows:
\begin{equation}
\tilde R_i(E)\equiv 
\int\limits_{\displaystyle{E^\prime_{i1}}}^{\displaystyle{E^\prime_{i2}}}
 R(E^\prime,E){\rm d}E^\prime,
\end{equation}
we obtain
\begin{equation}
S_i = \int\limits_0^{\infty}  f(E) \tilde R_i(E) {\rm d}E.
\label{eqn:si_int}
\end{equation}
From the above equation we see that as long as we are interested in the observed
count rate $S_i$ of a given data channel $i$, we get that number by integrating
the model spectrum multiplied by the effective area $\tilde R_i(E)$ for that
particular data channel. We have approximated $f(E)$ for each model bin $j$ by
(\ref{eqn:f1}), so that (\ref{eqn:si_int}) becomes
\begin{equation}
S_i = \sum\limits_j^{} F_j \tilde R_i(E_{a,j}),
\label{eqn:sisum}
\end{equation}
where as before $E_{a,j}$ is the average energy of the photons in bin $j$ given
by (\ref{eqn:eq}), and $F_j$ is the total photon flux for bin $j$, in e.g.
photons\,m$^{-2}$\,s$^{-1}$. It is seen from (\ref{eqn:sisum}) that we need to
evaluate $\tilde R_i$ not at the bin centre $E_j$ but at $E_{a,j}$, as expected.

Formally we may split up $\tilde R_i(E)$ in an effective area part $A(E)$ and a
redistribution part $\tilde r_i(E)$ in such a way that
\begin{equation}
\tilde R_i(E) = A(E) \tilde r_i(E).
\end{equation}
We have chosen our binning already in such a way that we have sufficient
accuracy when the total effective area $A(E)$ within each model energy grid bin
$j$ is approximated by a linear function of the photon energy $E$. Hence the
arf-part of $\tilde R_i$ is of no concern.  We only need to check how the
redistribution (rmf) part $\tilde r_i$ can be calculated with sufficiently
accuracy.  

For $\tilde r_i$ the arguments are exactly the same as for $A(E)$ in the sense
that if we approximate it locally for each bin $j$ by a linear function of
energy, the maximum error that we make is proportional to the second derivative
with respect to $E$ of $\tilde r_i$, cf. (\ref{eqn:epsilon_max}).

In fact, for a Gaussian redistribution function the following is straightforward
to prove:

\begin{theorem}
Assume that for a given model energy bin $j$ all photons are located at the
upper bin boundary $E_j+\Delta /2$. Suppose that for all data channels we
approximate $\tilde r_i$ by a linear function of $E$, and the coefficients are
the first two terms in the Taylor expansion around the bin centre $E_j$. Then
the maximum error $\delta$ made in the cumulative count distribution (as a
function of the data channel) is given by (\ref{eqn:de1}) in the limit of small
$\Delta$.
\label{the:rmfacc}
\end{theorem}

The importance of the above theorem is that it shows that the binning for the
model energy grid that we have chosen in Sect.~\ref{sect:modelbinning} is also
sufficiently accurate so that $\tilde r_i(E)$ can be approximated by a linear
function of energy within a model energy bin $j$, for each data channel $i$. 
Since we already showed that our binning is also sufficient for a similar linear
approximation to $A(E)$, it also follows that the total response $\tilde R_i(E)$
can be approximated by a linear function. Hence, within the bin $j$ we use
\begin{equation}
\tilde R_i(E_{a,j}) = \tilde R_i(E_j) + 
\frac{{\rm d}\tilde R_i }{ {\rm d}E_j}(E_j)\,\, (E_{a,j} - E_j).
\end{equation}

Inserting the above in (\ref{eqn:sisum}) and comparing with (\ref{eqn:rmatrix})
for the classical response matrix, we finally obtain 
\begin{equation}
S_i = \sum\limits_{j}^{} R_{ij} F_j + R^\prime_{ij} (E_{a,j} - E_j) F_j,
\label{eqn:rmatrixder}
\end{equation}
where $R_{ij}$ is the classical response matrix, evaluated for photons at the
bin centre, and $R^\prime_{ij}$ is its derivative with respect to the photon
energy $E_j$. In addition to the classical convolution, we thus get a second
term containing the relative offset of the photons with respect to the bin
centre. This is exactly what we intended to have when we argued that the number
of bins could be reduced considerably by just taking that offset into account.
It is just at the expense of an additional derivative matrix, which means only a
factor of two more storage space and computation time. But for this extra
expenditure we gained much more storage space and computational efficiency
because the number of model bins is reduced by a factor between 10--100.

Finally we make a practical note. The derivative $R^\prime_{ij}$ can be
calculated in practice either analytically or by numerical differentiation. In
the last case, it is more accurate to evaluate the derivative by taking the
difference at $E_j+\Delta /2$ and $E_j-\Delta /2$, and, wherever possible, not
to evaluate it at one of these boundaries and the bin centre. This last
situation is perhaps only unavoidable at the first and last energy value.

Also, negative response values should be avoided. Thus it should be ensured that
$R_{ij} + R^\prime_{ij} h$ is  non-negative everywhere for $-\Delta /2\le h \le
\Delta /2$. This can be translated into the constraint that $R^\prime_{ij}$
should be limited always to the following interval:
\begin{equation}
-2R_{ij}/\Delta E \le  R^\prime_{ij} \le 2R_{ij}/\Delta E.
\label{eqn:rdercrit}
\end{equation}
Whenever the calculated value of $R^\prime_{ij}$ should exceed the above limits,
the limiting value should be inserted instead. This situation may happen, for
example for a Gaussian redistribution for responses a few $\sigma$ away from the
centre, where the response falls off exponentially. However, the response
$R_{ij}$ is small for those energies anyway, so this limitation is not serious;
this is only because we want to avoid negative predicted count rates.

\section{Constructing the response matrix\label{sect:construct}}

In the previous section we outlined the basic structure of the response matrix.
For an accurate description of the spectrum, we need both the response matrix
and its derivative with respect to the photon energy. We build on this to
construct general response matrices for more complex situations.

\subsection{Dividing the response into components}

Usually only the non-zero matrix elements of a response matrix are stored and
used. This is done to save both storage space and computational time. The
procedure as used in XSPEC \citep{arnaud1996} and the older versions of SPEX
\citep{kaastra1996} is  that for each model energy bin $j$ the relevant column
of the response matrix is subdivided into groups. A group is a contiguous piece
of the column with non-zero response. These groups are stored in a specific
order; starting from the lowest energy, all groups belonging to a single photon
energy are stored before turning to the next higher photon energy.

This is not optimal neither in terms of storage space nor computational
efficiency, as illustrated by the following example. For the XMM-Newton/RGS, the
response consists of a narrow Gaussian-like core with a broad scattering
component due to the gratings in addition. The FWHM of the scattering component
is $\sim$10 times broader than the core of the response. As a result, if the
response is saved as a classical matrix, we end up with one response group per
energy, namely the combined core and wings response because they overlap in the
region of the Gaussian-like core. As a result, the response file becomes large.
This is not necessary because the scattering contribution with its ten times
larger width needs to be specified only on a model energy grid with ten times
fewer bins, compared to the Gaussian-like core. Thus, by separating out the core
and the scattering contribution, the total size of the response matrix can be
reduced by about a factor of 10. Of course, as a consequence each contribution
needs to carry its own model energy grid with it.

Therefore we propose  subdividing the response matrix into its constituent
components. Then for each response component, the optimal model energy grid can
be determined according to the methods described in
Sect.~\ref{sect:modelbinning}, and this model energy grid for the component can
be stored together with the response matrix part of that component. 
Furthermore, at any energy each component may have at most one response group.
If there were more response groups, the component should be subdivided further.

In X-ray detectors other than the RGS detector the subdivision could be
different. For example, with CCD detectors one could split up the response into
for components: the main diagonal, the Si fluorescence line, the escape peak,
and a broad component due to split events.

\subsection{Complex configurations}

In most cases an observer analyses the data of a single source with a single
spectrum and response for a single instrument. However, more complicated
situations may arise. Examples are:

\begin{enumerate}

\item A spatially extended source, such as a cluster of galaxies, with observed
spectra extracted from different regions of the detector, but with the need to
be analysed simultaneously due to the overlap in point-spread function from one
region to the other. 

\item For the RGS of XMM-Newton, the actual data space in the dispersion
direction is actually two-dimensional: the position $z$ of a photon on the
detector and its energy or pulse height $E^\prime$ measured with the CCD
detector. Spatially extended X-ray sources are characterised by spectra that are
a function of both the energy $E$ and off-axis angle $\phi$. The sky photon
distribution as a function of $(\phi,E)$ is then mapped onto the
$(z,E^\prime)$-plane. One may analyse such sources by defining appropriate
regions in both planes and evaluating the correct (overlapping) responses.

\item One may also fit simultaneously several time-dependent spectra using the
same response, for example data obtained during a stellar flare.

\end{enumerate}

It is relatively easy to model all these situations (provided that the
instrument is understood sufficiently, of course), as we show below.

\subsection{Sky sectors}

First, the relevant part of the sky is subdivided into sectors, each sector
corresponding to a particular region of the source, for example a circular
annulus centred around the core of a cluster or an arbitrarily shaped piece of a
supernova remnant, etc.

A sector may also be a point-like region on the sky. For example if there is a
bright point source superimposed upon the diffuse emission of the cluster, we
can define two sectors:  an extended sector for the cluster emission and a
point-like sector for the point source.  Both sectors might even overlap, as
this example shows.  

Another example is the two nearby components of the close binary $\alpha$~Cen
observed with the XMM-Newton instruments with overlapping psfs of both
components.  In that case we would have two point-like sky sectors, each sector
corresponding to one of the double star components.

The model spectrum for each sky sector may and is different in general. For
example, in the case of an AGN superimposed upon a cluster of galaxies, one
might model the spectrum of the point-like AGN sector with a power law and the
spectrum from the surrounding cluster emission with a thermal plasma model.

\subsection{Detector regions}

The observed count rate spectra are extracted in practice in different  regions
of the detector. It is necessary to distinguish  the (sky) sectors and
(detector) regions clearly. A detector region for the XMM-Newton EPIC camera
would be for example a rectangular box, spanning a certain number of pixels in
the x- and y-directions. It may also be a circular or annular extraction region
centred around a particular pixel of the detector, or whatever spatial filter is
desired. 

The detector regions need not coincide with the sky sectors and  their number
should not be equal. A good example of this is again an AGN superimposed upon a
cluster of galaxies.  The sky sector corresponding to the AGN is simply a point,
while, for a finite instrumental psf, its extraction region at the detector is
for example a circular region centred around the pixel corresponding to the sky
position of the AGN.

Also, one could observe the same source with a number of different instruments
and analyse the data simultaneously. In this case one would have only one sky
sector but more detector regions, namely one for each participating instrument.

\subsection{Consequences for the response}

In all cases mentioned above, with more than one sky sector or more than one
detector region involved, the response contribution for each combination of sky
sector - detector region must be generated. In the spectral analysis,  the model
photon spectrum is calculated for each sky sector, and all these model spectra
are convolved with the relevant response contributions  to predict the count
spectra for all detector regions. Each response contribution for a sky sector -
detector region combination itself may consist again of different response
components, as outlined in the previous subsection.

Combining all this, the total response matrix then consists of a list of
components, each component corresponding to a particular sky sector and detector
region.  For example, we assume that the RGS has two response contributions: one
corresponding to the core of the lsf and the other to the scattering wings. We
assume that this instrument observes a triple star where the instrument cannot
resolve two of the three components.  The total response for this configuration
then consists of 12 components. These components include three sky sectors,
assuming each star has its own characteristic spectrum, times two detector
regions, including a spectrum extracted around the two unresolved stars and one
around the other star, times two instrumental contributions (the lsf core and
scattering wings).

\section{Proposed file formats}\label{sect:formats}

In the previous sections it was shown how the optimal data and model binning can
be determined, and the corresponding optimal way to create the instrumental
response. Now we focus on the possible data formats for these files.

A widely used response matrix format is NASA's OGIP\footnote{see the OGIP manual
by K.A. Arnaud and I.M. George at
\url{https://heasarc.gsfc.nasa.gov/docs/heasarc/ofwg/docs/spectra/ogip_92_007/ogip_92_007.html}}
\citep{corcoran1995} FITS format \citep{wells1981}. This is used, for example,
as the data format for XSPEC. There are a few reasons that we propose not to
adhere to the OGIP format in its present form here, as listed below:

\begin{enumerate}

\item The OGIP response file format, as it is currently defined, does not
account for the possibility of response derivatives. As was shown in previous
sections, these derivatives are needed for the optimal binning. 

\item As was shown in this work, it is more efficient to do the grouping within
the response matrix differently, splitting the matrix into components where each
component may have its own energy grid. This is not possible within the present
OGIP format.

\end{enumerate}

 The FITS format used by the SPEX package version 2 obeys all the constraints
that we outlined in this paper \citep{kaastra1996}. This format is described in
detail in the manual of that package.\footnote{See \url{www.sron.nl/spex}}

\section{Conclusion}

In this paper, we derived  the optimal bin size for both binning X-ray data as
well as binning the model energy grids used to calculate predicted X-ray
spectra. 

The optimal bin size for model energy grids, in the way it is usually used (i.e.
putting all photons at the bin centres) requires a large number of model bins.
We have shown here that the number of model bins can be reduced by an order of
magnitude or more using a variable binning scheme where not only the number of
photons but also their average energy within a bin are calculated. The basic
equations are (\ref{eqn:app1}), (\ref{eqn:de1ar}), and (\ref{eqn:de1tot}).

The data binning depends mostly on the number of counts per resolution element,
and to a lesser extent on the number of resolution elements. The commonly used
prescription of binning to 1/3 of the instrumental FWHM is a safe number to use
for most practical cases, but in most cases a somewhat courser sampling can be
used without losing any sensitivity to test spectral models. The recommended
binning scheme is given by (\ref{eqn:databin}).

Correspondingly, the response matrix needs to be extended to two components: the
`classical' part and its derivative with respect to photon energy (See
eq.~(\ref{eqn:rmatrixder})). With these combined improvements a strong reduction
of data storage needs and computational time is reached, which is very useful
for the analysis of high-resolution spectra with large numbers of bins and
computational complex spectral models.

Finally, we have outlined a few simple methods to reduce the size of the
response matrix by splitting it into constituent components with different
spectral resolution.

All these improvements are fully available in the spectral fitting package SPEX.

\begin{acknowledgements}

SRON is supported financially by NWO, the Netherlands Organization for
Scientific Research. 

\end{acknowledgements}

\bibliographystyle{aa}
\bibliography{newbin}

\begin{appendix}

\section{Testing models\label{sect:testing}}

\subsection{How is it possible to test an approximation to the spectrum?\label{sect:optimal}}

\begin{figure}[!htbp]
\resizebox{\hsize}{!}{\includegraphics[angle=-90]{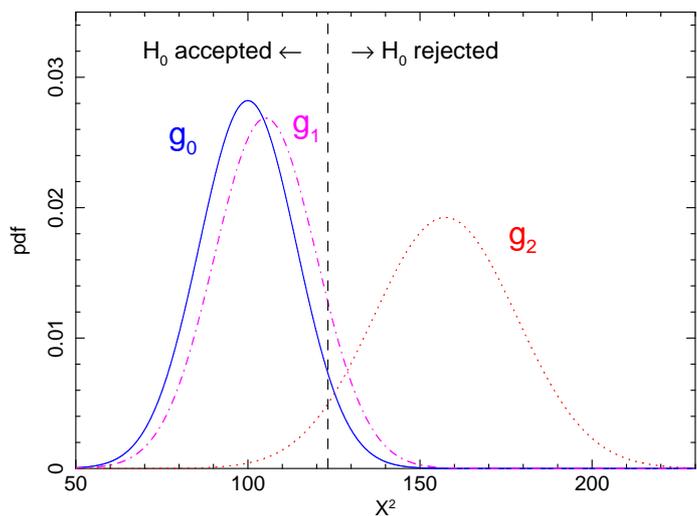}}
\caption{Probability distributions $g_i(T)$ for the example of $T=\chi^2$,
corresponding to different spectral distribution functions $f_0$ (the ``true''
model underlying the data), $f_1$ (an approximation to $f_0$ based on finite
sampling distance $\Delta$), and $f_2$ (an alternative physical model of the
spectrum). The calculation is made for $100$ degrees of freedom and
$\alpha=\beta=0.025$. Thus, 2.5\% of the area under $g_0$ is to the right of the
dashed vertical line, and 2.5\% of the area under $g_2$ is to the left of this
same line. Further, 5\% of the area under $g_1$ is to the right of the vertical
line.}
\label{fig:pow}
\end{figure}

We make use of probability distributions to assess the accuracy of an
approximation to the model spectrum. 

Suppose that $f_0(x)$ is the unknown, true probability density function (pdf)
describing the observed spectrum. Further let $f_1(x)$ be an approximation to
$f_0$ (discretely sampled with a bin size $\Delta$).  Further let $f_2(x)$ be
the pdf of an alternative spectral model. Ideally, one now designs a statistical
test, based on some test statistics $T$. Choices for $T$ would be $\chi^2$ in a
chi-squared test, or $D$ (see later) in a Kolmogorov-Smirnov test. The
hypothesis $H_0$ that $f_0(x)$ describes the spectrum is accepted if $T$ is
smaller than some threshold $c_\alpha$, and rejected if $T>c_\alpha$.

The probability $\alpha$ that $H_0$ will be rejected if it is true is called the
size of the test. The probability that $H_0$ will be rejected if $H_2$ is true
is called the power of the test and is denoted by $1-\beta$ as follows:
\begin{eqnarray}
{\rm Prob}(T>c_\alpha \vert H_0) & = & \alpha,\nonumber \label{eqn:crith0h2}\\
{\rm Prob}(T>c_\alpha \vert H_2) & = & 1-\beta. 
\end{eqnarray}
An ideal statistical test of $H_0$ against $H_2$ has $\beta$ small (large
power). However, in our case we are not interested in getting the most
discriminative power between the alternatives $H_0$ and $H_2$, but we want to
know how well the approximation $f_1$ works in reaching the right conclusions
about the model. When we use $f_1$ instead of $f_0$, we want to reach the same
conclusion in the majority of all cases. We define the size of the test of $H_1$
versus $H_2$ to be $k\alpha$ with the same $\alpha$ as above. For $k=1$, both
distributions would of course be equal, but this happens only for bin size
$\Delta=0$.

We adopt as a working value $k=2$, $\alpha=0.025$. That is, using a classical
$\chi^2$ test $H_1$ will be rejected against $H_2$ in 5\% of the cases, the
usual criterion in $\chi^2$-analysis. This means that in practice $H_1$ reaches
the same correct conclusion as $H_0$ in $0.95/0.975\simeq 0.975$ of all cases.
For an example, see Fig.~\ref{fig:pow}.

\subsection{Statistical tests}

Up to now we have not specified the test statistic $T$. It is common to do a
$\chi^2$-test. However, the $\chi^2$-test has some drawbacks, in particular for
small sample sizes. It can be shown, for instance, for a Gaussian distribution
that the test is rather insensitive in discriminating relatively large relative
differences in the tails of the distribution. 

For our purpose, a Kolmogorov-Smirnov test is more appropriate. This powerful,
non-parametric test is based upon the test statistic $D$ given by
\begin{equation}
\label{eqn:kolstat}
D = \max \vert S_N(x) - F(x) \vert,
\end{equation}
where $S_N(x)$ is the observed cumulative distribution for the sample of size
$N$, and $F(x)$ is the cumulative distribution for the model that is tested. 
Clearly, if $D$ is large, $F(x)$ is a bad representation of the observed data
set. The statistic $D^\prime\equiv\sqrt{N}D$ for large $N$ (typically 10 or
larger) has the limiting Kolmogorov-Smirnov distribution, with expectation value
$\sqrt{\pi/2}\ln 2$=0.86873 and standard deviation $\pi\sqrt{1/12-(\ln
2)^2/2\pi}$=0.26033. The hypothesis that the true cumulative distribution is $F$
will be rejected if $D^\prime >c_\alpha$, where the critical value $c_\alpha$
corresponds to a certain size $\alpha$ of the test.  

A good criterion for determining the optimal bin size $\Delta$ is that the
maximum difference
\begin{equation} 
\delta_k \equiv  \ \max_m^{} \vert F_1(m\Delta) - F_0(m\Delta) \vert
\label{eqn:lambda_k}
\end{equation}
should be sufficiently small compared to the statistical fluctuations described
by $D^\prime$. Here $F_0(x)$ is the true underlying cumulative distribution and
$F_1(x)$ is again its approximation, which we assume here are sampled on an
energy grid with spacing $\Delta$.

We can apply exactly the same arguments as discussed before and use
Fig.~\ref{fig:pow} with $T=D^\prime$.

The Kolmogorov-Smirnov test was designed for comparing distributions based on
drawing $N$ independent values from the distribution $f(x)$; it is assumed that
$x$ can have any real value within the interval where $f(x)>0$. In our case,
because of the binning with finite bin size $\Delta$, we first round the
observed values of $x$ to a set of discrete values and then perform a test. This
makes a significant difference; for a binned Gaussian redistribution function
$f(x)$ with $N$ large, we found from numerical simulations that the average
value of $D^\prime$ equals 0.64, 0.69, 0.77, and 0.84 for $\Delta$ equal to 0.5,
0.3, 0.1, and 0.01, respectively. Thus, convergence to the asymptotic value for
$D^\prime$ of 0.86873 for $\Delta\rightarrow 0$ is rather slow.

\subsection{Analytical approximation\label{sect:anap}}

We now make a simple analytical approximation for the optimal bin size. We first
determine the maximum deviation of the approximation $F_1$ from $F_0$: 
$\delta_k$ given by (\ref{eqn:lambda_k}). This maximum occurs at a given value
of $x_m$. We are interested in tests of $H_0$ versus alternatives $H_2$, and as
we have seen (Fig.~\ref{fig:pow}), this requires knowledge of the distribution
$g_0(D^\prime)$ at high values of $D^\prime$ (typically the highest 5\% of the
distribution). Because the observed sample is drawn from the distribution
$f_0(x)$, $D^\prime$ can be reached at any value of $x$, wherever the random
fluctuations reach the highest values. However, in the test using $f_1$ as an
approximation to $f_0$, the $x$-values near $x_m$ have a higher probability of
yielding the maximum because in addition to the random noise there is an offset
$\lambda_k$ due to the maximum difference between $F_0$ and $F_1$. In fact, we
may approximate the tail of the distributions $g_0$ and $g_1$ by
\begin{equation}
g_1(x+\lambda_k)\simeq g_0(x).
\label{eqn:ks-shift}
\end{equation}
We now approximate the distribution function $g_0$ of the $D^\prime$ statistic
by the limiting Kolmogorov-Smirnov distribution $g_{\rm{KS}}$. We thus have to
solve for
\begin{eqnarray}
 G_0(c_\alpha)  &=& 1-\alpha, \\
 G_1(c_\alpha)  &=& 1-k\alpha,
\end{eqnarray}
which translates into
\begin{eqnarray}
\label{eqn:ksh0r}
 G_{\rm{KS}}(c_\alpha)  &=& 1-\alpha, \\
\label{eqn:ksh1r}
 G_{\rm{KS}}(c_\alpha-\lambda_k)  &=& 1-k\alpha.
\end{eqnarray}
For our choice of $\alpha=0.025$ and $k=2,$ we obtain $\lambda_k = 0.122$.  We note
that the dependence on $\alpha$ is weak: for $\alpha=0.01$ or $\alpha=0.05,$ we
get $\lambda_k$ values of 0.110 and 0.134, respectively. To get the optimal bin
size for any value of $N$, we calculate 
\begin{equation}
\epsilon \equiv \lambda_k / \sqrt{N}.
\label{eqn:epsilon_lambda}
\end{equation}
We then approximate $\delta_k\simeq \epsilon$ and using (\ref{eqn:lambda_k}) we
can determine for which bin size $\Delta$ this occurs.

In evaluating (\ref{eqn:lambda_k}) for a given cumulative distribution $F_0(x)$
and $F_1(x)$, it is useful to remember that the extrema of $F_1(x)-F_0(x)$ occur
where its derivative is zero, i.e. where $f_1(x) = f_0(x),$ where $f_0$ and
$f_1$ are the corresponding probability density distributions.

\subsection{Monte Carlo calculations\label{sect:mccalculations}}

The analytical approximations described above yield interesting asymptotic
limits. However, we need to test their accuracy and applicability. To do this,
we have used Monte Carlo simulations. For a set of sample sizes $N$ and bin
sizes $\Delta$, we  calculated the true distribution function $F_0$ and its
approximation $F_1$. Using $F_0$, we have drawn a large number (typically
$10^7$) of realisations. For each random realisation, we calculated $D^\prime$
against both models $F_0$ and $F_1$. Combining all runs, we determined the
cumulative distributions $G_0(D^\prime)$ and $G_1(D^\prime)$. From these
distributions, we determined the critical value $c_\alpha$ where $G_0(c_\alpha)
= 1-\alpha$ (using $\alpha=0.025$). For a given sample size $N$, we then varied
$\Delta$ in such a way that $G_1(c_\alpha) = 1-k\alpha$ with as before $k=2$.
This procedure then yields the optimal bin size $\Delta$ as a function of $N$.

The Monte Carlo method works well for large values for $N$. Because we work with
$\alpha=0.025$, for $N\leq 1/\alpha$ the distribution $G_0(x)$ begins suffering
from discontinuous steps due to the limited number $N$ of values that the
observed distribution $S_N(x)$ can have. In practice, this is not a major
problem as our results to be presented later show that in those cases the
derived bin sizes become of the order of the FWHM of the instrumental response
function anyway.

\subsection{Extension to complex spectra and instruments}

The estimates that we derived for determining bin sizes hold for any spectrum.
The optimal bin size $\Delta$ as a function of $N$ can be determined using the
analytical approximations or numerical calculations outlined above.

This method is not always practical, however. First, the spectral resolution can
be a function of the energy, hence binning with a uniform bin width over the
entire spectrum is not always desired.

Further, regions with poor statistics contain less information than regions with
good statistics, hence can be sampled on a much coarser grid.

Finally, a spectrum often extends over a much wider energy band than the
spectral resolution of the detector. In order to estimate $\epsilon$, this would
require that first the model spectrum, convolved with the instrument response,
should be used to determine $f_0(x)$. However, the true model spectrum is known
in general only after spectral fitting. But the spectral fitting can be done
only after a bin size $\Delta$ has been set.

To overcome these problems, we consider the instrumental line-spread function
(lsf). This is the response (probability distribution) of mono-energetic photons
that the instrument measures. Most X-ray spectrometers have a lsf with a
characteristic width $\Delta E$ smaller than the incoming photon energy $E$, and
we only consider  instruments obeying that condition; there is little use in
binning data from instruments lacking spectral resolution. To be more precise,
we define a resolution element to be a region in data space corresponding to
the  FWHM of the detector; this FWHM is called $\Delta E$.

We now consider a region $r$ of the spectrum with a width of the order of
$\Delta E$. In this region the observed spectrum is mainly determined by the
convolution of the instrumental lsf with the true model spectrum in an energy
band with a width of the order of $\Delta E$. We ignore here any substantial
tails in the lsf, since they are of little importance in the determination of
the optimal bin size/resolution.

If all $N_r$ photons within the resolution element $r$ would be emitted at the
same energy $E_r$, and no photons would be emitted at any other energy, the
observed spectrum would be simply the lsf for energy $E_r$ multiplied by the
number of photons $N_r$. For this situation, we can easily evaluate $\lambda_k$
for a given binning. 

If the spectrum is not mono-energetic, then the observed spectrum within the
resolution element is the convolution of the full photon distribution with the
lsf. In this case, the observed spectrum $S_r$ in resolution element $r$ is a
weighted sum of the photon spectra $F_i$ in a number of usually adjacent
resolution elements $i$, say $S_r = \sum w_i F_i$. Because for a set of weights
$w_i$ in general $\sqrt{\sum w_i^2} \le \sum w_i$, the squared modelling
uncertainties $\epsilon_r^2$ are smaller for the case of a broadband spectrum
than for a mono-energetic spectrum. It follows that if we determine
$\epsilon_r^2$ for the mono-energetic case from the sampling errors in the lsf,
this is an upper limit to the true value for $\epsilon^2$ within the resolution
element $r$.

We now need to combine the resolution elements. Suppose that in each of the $R$
resolution elements we perform a Kolmogorov-Smirnov test. For the maximum of
$D^\prime$ over all resolution elements $r$, we test 
\begin{equation}
\label{eqn:maxd}
\delta \equiv \max_r \sqrt{N_r}D_r,
\end{equation}
where as before $N_r$ is the number of photons in the resolution element $r$,
and $D_r$ is given by (\ref{eqn:kolstat}) over the interval $r$. All the
$\sqrt{N_r}D_r$ values are independently distributed, hence the cumulative
distribution function for their maximum is simply the $R$th power of a single
stochastic variable with the relevant distribution. Therefore we obtain
\begin{eqnarray}
\label{eqn:ksh0rb}
[ G_0(c_\alpha) ]^R &=& 1-\alpha, \\
\label{eqn:ksh1rb}
[ G_1(c_\alpha) ]^R &=& 1-k\alpha.
\end{eqnarray}

\section{Determining the grids in practice}

Based upon the theory developed in the previous sections, we present here a
practical set of algorithms for the determination of both the optimal data
binning and the model energy grid determination. This may be helpful for the
practical purpose of developing software for a particular instrument to
construct the relevant response matrix.

\subsection{Creating the data bins}

Given an observed spectrum obtained by some instrument, the following steps
should be performed to generate an optimally binned spectrum.

\begin{enumerate}

\item Determine for each original data channel $i$ the nominal energy $E_{j0}$,
defined as the energy for which the response at channel $i$ reaches its maximum
value. In most cases, this is the nominal channel energy.

\item Determine for each data channel $i$ the limiting points $(i1,i2)$ for the
FWHM in such a way that $R_{k,j0}\ge 0.5\,R_{i,j0}$ for all $i1\le k\le i2,$
while the range of $(i1,i2)$ is as broad as possible.

\item By (linear) interpolation, determine for each data channel the points
(fractional channel numbers) $c1$ and $c2$ near $i1$ and $i2,$ where the
response is actually half its maximum value. By virtue of the previous step, the
absolute difference $\vert c1-i1\vert$ and $\vert c2-i2\vert$ never can exceed
1.

\item Determine for each data channel $i$ the FWHM $c_i$ in units of channels,
by calculating $c2-c1$. Assure that $c_i$ is at least 1.

\item Determine for each original data channel $i$ the FWHM in energy units
(e.g. in keV). Call this $W_i$. This and the previous steps may of course also
be performed directly using instrument calibration data.

\item Determine the number of resolution elements $R$ by the following
approximation:

\begin{equation}
R = \sum\limits_i^{} \frac{1}{ c_i}. 
\label{eqn:rdet}
\end{equation}

\item Determine for each bin the effective number of events $N_r$ from the
following expressions:

\begin{eqnarray}
C_r &=& \sum\limits_{k=i1-1}^{i2+1} C_k, \\
h_r &=& \sum\limits_{k=1}^{N_c} R_{k,j0} / 
        \sum\limits_{k=i1-1}^{i2+1} R_{k,j0}, \\
N_r &=& C_r h_r.
\label{eqn:nrcalc}
\end{eqnarray}

In the above, $C_k$ is the number of observed counts in channel $k$, and $N_c$
is the total number of channels.  Take care that in the summations $i1-1$ and
$i2+1$ are not out of their valid range $(1,N_c)$. If for some reason there is
not a first-order approximation available for the response matrix $R_{k,j}$ then
one might simply approximate $h_r$ from e.g. the Gaussian approximation, namely
$h_r = 1.314$; cf. Sect.~\ref{sect:datafinal}. This is justified since the
optimal bin size is not a strong function of $N_r$; cf. Fig.~\ref{fig:databin}.
Even a factor of two error in $N_r$ in most cases does not affect the optimal
binning too much.

\item Using (\ref{eqn:databin}), determine for each data channel the optimal
data bin size in terms of the FWHM. The true bin size $b_i$ in terms of number
of data channels is obtained by multiplying this by $c_i$ calculated above
during step~4.  Make $b_i$ an integer number by ignoring all decimals (rounding
it to below), but take care that $b_i$ is at least 1.

\item It is now time to merge the data channels into bins. In a loop over all
data channels, start with the first data channel. Name the current channel $i$.
Take in principle all channels $k$ from channel $i$ to $i+b_i-1$ together.
However, check that the bin size does not decrease significantly over the
rebinning range. In order to do that check, determine for all $k$ between $i$
and $i+b_i-1$ the minimum $a_i$ of $k+b_k$. Extend the summation only from
channel $i$ to $a_i-1$. In the next step of the merging, $a_i$ becomes the new
starting value $i$. The process is finished when $a_i$ exceeds $N_c$.

\end{enumerate}

\subsection{Creating the model bins}

After having created the data bins, it is possible to generate the model energy
bins. Some of the information obtained from the previous steps that created the
data bins is needed.

The following steps need to be taken:

\begin{enumerate}

\item Sort the FWHM of the data bins in energy units ($W_i$) as a function of
the corresponding energies $E_{j0}$. Use this array to interpolate any true FWHM
later. Also use the corresponding values of $N_r$ derived during that same
stage. Alternatively, one may use directly the FWHM as obtained from calibration
files.

\item Choose an appropriate start and end energy, e.g. the nominal lower and
upper energy of the first and last data bin with an offset of a few FWHMs (for a
Gaussian, about 3 FWHM is sufficient). In the case of a lsf with broad wings
(like the scattering due to the RGS gratings), it may be necessary to take an
even broader energy range.

\item In a loop over all energies, as determined in the previous steps,
calculate the bin size in units of the FWHM using (\ref{eqn:app1}).

\item Also, determine the effective area factor $\frac{{\rm d}\ln E }{{\rm d}\ln
A}$ for each energy; one may do that using a linear approximation.

\item For the same energies, determine the necessary bin width in units of the
FWHM using eqn.~(\ref{eqn:de1tot}). Combining this with the FWHMs determined
above gives for these energies the optimal model bin size $\Delta E$ in keV.

\item Now the final energy grid can be created. Start at the lowest energy
$E_{1,1}$, and interpolate in the $\Delta E$ table the appropriate $\Delta E
(E_{1,1})$ value for the current energy.  The upper bin boundary $E_{2,1}$ of
the first bin is then simply $E_{1,1}+\Delta E(E_{1,1})$.

\item Using the recursive scheme $E_{1,j} = E_{2,j-1}$, $E_{2,j} = E_{1,j} +
\Delta E(E_{1,j})$ determine all bin boundaries until the maximum energy has
been reached.  The bin centres are simply defined as $E_j =
0.5(E_{1,j}+E_{2,j})$.

\item Finally, if there are any sharp edges in the effective area of the
instrument, it is necessary to add these edges to the list of bin boundaries.
All edges should coincide with bin boundaries.

\end{enumerate}

\section{Oversampling of data\label{sect:appc}}

If a spectrum is binned with  data bins that are too narrow, important
information about the spectrum is lost. This was shown in a more formal way, for
instance by \citet[sect.~2.6]{kaastra1999} using $\chi^2$-statistics. The
argument also holds for lower count statistics. Here we give a practical example
using C-statistics, but our conclusions are independent of the precise
statistics used.

\begin{figure}[!htbp]
\resizebox{\hsize}{!}{\includegraphics[angle=-90]{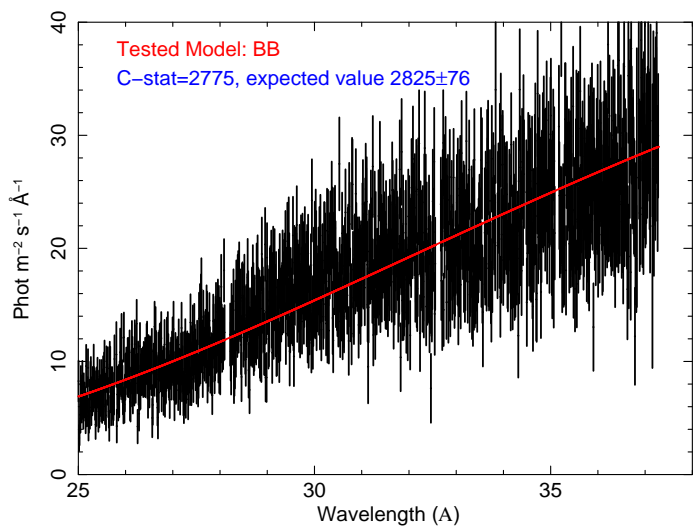}}
\caption{Simulated X-ray spectrum (data points) sampled with a bin size of
0.01~\AA\ and fitted with a blackbody with a temperature of 50~eV (solid line).
The goodness of the fit is indicated in the figure.}
\label{fig:chibb1}
\end{figure}

We have simulated a spectrum consisting of a 50~eV blackbody plus a weak
Gaussian emission line at 30~\AA\ with a FWHM of 1~\AA. The instrument
over-samples the spectrum with a bin size of 0.01~\AA. The spectrum was fit with
a simple blackbody (without a Gaussian line) using C-statistics and the fit was
'perfect', with a C-stat value of 2775, to be compared to the expected value of
$2825\pm 76$ for the best-fit model. See Fig.~\ref{fig:chibb1} for the fit. For
the fitting we have used the SPEX software, which calculates for each fit the
expected value of C-stat (in this case 2825) and its variance. This expected
value is simple to calculated knowing the expected number of counts from the
best-fit model and the observed number of counts from the data in each data
channel.

\begin{figure}[!htbp]
\resizebox{\hsize}{!}{\includegraphics[angle=-90]{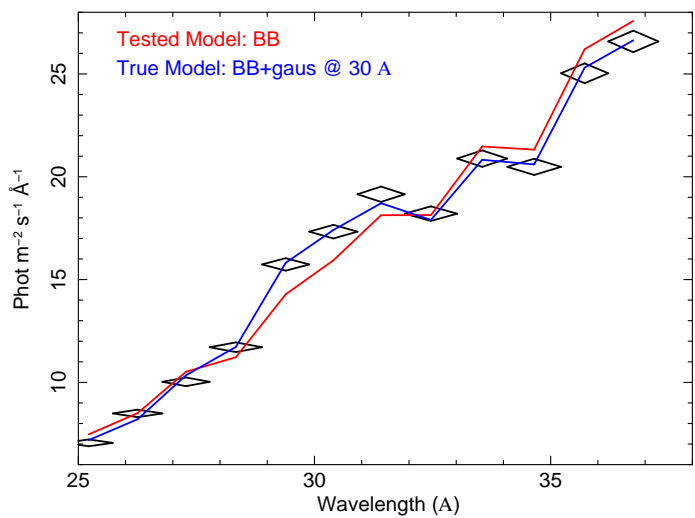}}
\caption{Same data as Fig.~\ref{fig:chibb1} but rebinned by a factor of 100
(diamonds). The same best-fit model of Fig.~\ref{fig:chibb1} using a simple
blackbody with a temperature of 50~eV is indicated with the red line. The blue
line indicates the model for the source that entered the simulation: the sum of
a blackbody spectrum and a weak Gaussian line at 30~\AA.}
\label{fig:chibb2}
\end{figure}

Thus, with this bin size, the observer would conclude that a simple blackbody
spectrum yields an accurate representation of the data, and there is no need to
add more components because the present fit is already statistically acceptable.
However, in Fig.~\ref{fig:chibb2} we show the same data and model rebinned by a
factor of 100 (i.e. 1~\AA\ wide bins). It is seen only at this bin size that the
source has an emission line in addition to the simple blackbody model.

\begin{figure}[!htbp]
\resizebox{\hsize}{!}{\includegraphics[angle=-90]{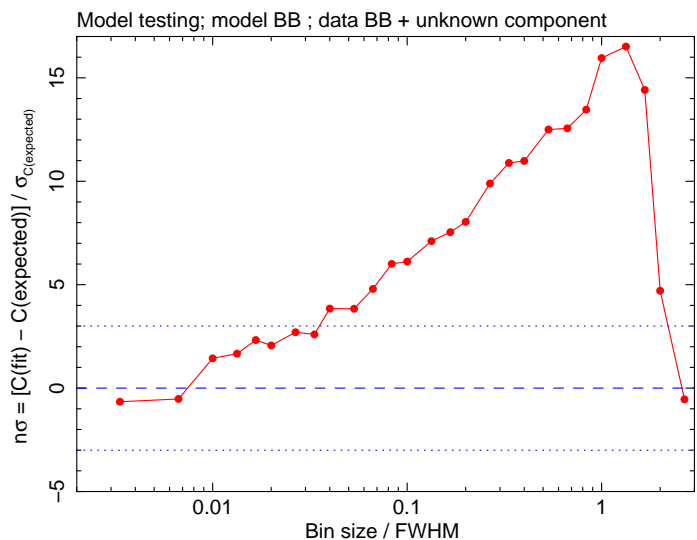}}
\caption{Best-fit C-statistic of the spectrum of Fig.~\ref{fig:chibb1} when
fitted with a simple blackbody. We subtract here the expected value of C-stat
and scale by its expected root-mean-square value, so that it is expressed in
equivalent number of $\sigma$. The dotted lines indicate the $\pm 3\sigma$
range; when the best fit is outside this $\pm 3\sigma$ range, most investigators would
reject the model (in this case a simple blackbody). The FWHM used in this figure
is the FWHM of the additional Gaussian component in the spectrum (1~\AA).}
\label{fig:chibb3}
\end{figure}

We elaborate on this by rebinning the original spectrum of Fig.~\ref{fig:chibb1}
by different factors, fitting the spectrum with a single blackbody component and
comparing the best-fit C-stat with the relevant expected C-stat and its
variance. The results are shown in Fig.~\ref{fig:chibb3}. It is clear that for
all bin sizes less than about 0.04~\AA\ the observer cannot reject the simple
blackbody hypothesis, and has no reason to add the Gaussian component to the
model. Only for bin sizes of the order of 1~\AA, the simple blackbody hypothesis
can be rejected at the $>15\sigma$ significance level. For higher binning
factors, the bin size becomes larger than the width of the Gaussian line, and
the significance drops again.

While the proper flux of the Gaussian line, including its uncertainties, is
recovered at the proper value, regardless of the binning, when the additional
Gaussian is introduced in the model, the detection and proof of existence of
this component is only achieved when the spectrum is optimally binned, in this
case (relatively weak line) with a bin size close to the FWHM of the Gaussian.

\end{appendix}

\end{document}